\documentclass[showpacs,preprintnumbers,prd,nofootinbib,floats,amssymb,floatfix]{revtex4}
\usepackage{amsmath}
\usepackage{amsfonts}
\usepackage{graphicx}
\usepackage{amssymb,latexsym}
\usepackage{hyperref}
\usepackage{amsmath}
\usepackage{amstext}
 
\begin{document}

\title{Hamiltonian dynamics and  Faddeev-Jackiw of the Jackiw-Teitelboim model written as a BF theory}
\author{J. Manuel-Cabrera} \email{jaime.manuel@ujat.mx}
\author{Jorge Mauricio Paulin Fuentes}
\affiliation{División Académica de Ciencias Básicas, Universidad Juárez Autónoma de Tabasco, Km 1 Carretera Cunduacán-Jalpa \\  Apartado Postal 86690, Cunduacán, Tabasco, México}

\begin{abstract}

The Jackiw-Teitelboim model written as a BF theory in two dimensions is analyzed by using Dirac's and Faddeev-Jackiw symplectic formalism. The analysis consists in finding the full structure of the constraints, the gauge transformations, the counting of degrees of freedom, and the generalized Faddeev-Jackiw brackets. The Poincar\'e symmetry and the diffeomorphisms are found. Further, we show that the Faddeev-Jackiw and  Dirac's brackets coincide with each other.
\end{abstract}

\date{\today}
\preprint{}
\maketitle
\section{INTRODUCTION}

The interest in studying lower dimensional theories is that lower dimensional models have been of enormous use in several branches of physics and specifically lower dimensional gravity theories have proven highly instructive in the understanding of quantum gravity theories. Also, they provide a starting point in which some physical phenomena can be easily demonstrated while circumventing the mathematical complexities often encountered in 4D. Besides, they help to generate new ideas and to stimulate new relations with other fields \cite{1,1a}. For instance, it is well-known the relation between  General Relativity in 3D and Chern-Simons action, developed by Witten in \cite{2}. In the context of  General Relativity in 2D, gravity 1+1 dimensions cannot be formulated on the Einstein-Hilbert action, therefore, it is necessary to invent a model. A 2D gravity model has been suggested by Jackiw y Teitelboim  \cite{2a, 3,4}. The connection betweenn Jackiw-Teitelboim JT  theory and BF formulation was analysed by  K. Isler and C.A.Kamimura \cite{Fu}, who also showed that  JT model can be written as a gauge theory. In \cite{Grumi}, they used the Jackiw-Teitelboim model in the linear dilaton sector considering different sets of AdS2 boundary conditions. On the other hand, S. Josephine argued in \cite{Josep} that it is possible to find  a general solution for correlators of external boundary operators in black holes states of JT model. In another recent paper \cite{Momeni}, the authors investigate the exact solutions of the JT model and the deformed JT, using all non-zero components of the dilatonic equations of motion by the appropriate integral transformation into the Euclidean time variable. Quite recently, D. Grummier, J. Hartong, S. Prohazka and J. Salzer has been studied various limits of JT gravity, including Newton-Cartan and Carrollian versions of dilaton gravity in two dimensions \cite{Gu}. As previously mentioned, the motivation to study gravity in low dimensions is that dimensional reduction is expected to allow suitable simplification to perform a complete analysis and preserve the relevant conceptual features of General Relativity in 4D.

This article aims to present two classical approaches to analyze the JT model written as a BF theory in 1+1 dimensions introduced by K. Isler and C.A.Kamimura \cite{Fu} by means of Dirac formalism and the Faddeev-Jackiw method. It is important to comment that there exist analysis of the BF action developed in a smaller phase space\footnote{This means that only those variables that occur in the action with temporal derivative are considered as dynamical \cite{Esca}.} through Dirac's algorithm reported in \cite{7,7a} and the Hamilton-Jacobi formalism has been studied in \cite{8}, however, in \cite{7, 7a}  the analysis was developed in a smaller phase space and the complete structure of the constraints on the full phase space was not reported. The implementation of Dirac's approach in the full phase space, it is important in the classification of the constraints in first and second class, in some cases, it is complicated and is not a trivial task \cite{9, 8a}, therefore, it is necessary to use another framework that could give us a complete description of singular systems. In this context, the Faddeev-Jackiw JT \cite{8b}  formalism with the Barcelos Neto-Wotzasek extension BW \cite{10a,10b} is an alternative scheme to the Dirac approach. The combination of JT and BW is sometimes  called in the literature as modified FJ formalism, or symplectic approach. We can find diverse applications, developments and the equivalence between Dirac method and the original JT in \cite{10a, 10b, 12a, 8bb, 8bbb, 10c, 10d}. Recently, the JT symplectic formalism has been employed to study the Bonzom-Livine action \cite{10e}, topologically massive gravity TMG \cite{10f}, and the abelian analog version of TMG theory at the chiral point \cite{10g} and also in four-dimensional General Relativity GR and to GR extensions \cite{10h}.

The advantages in using the JT approach come from the fact that it is not necessary to do the classification of the constraints in first and second class as carried out in Dirac algorithm since all the constraints of the theory are on the same footing. Additionally, all the relevant information of the theory can be obtained through an invertible symplectic matrix where the entries of this inverse matrix correspond to the JT generalized brackets and coincide with the Dirac brackets \cite{8bb, 8bbb}. Furtermore, Montani-Wotzasek MW \cite{12a}, presented the procedure of dealing with the gauge symmetry transformation over all the configuration space.

In this paper, we consider the analysis of JT theory in the BF formulation in the context of Dirac's formalism and JT symplectic approach. If we want to compare both scheme, it is mandatory to work in Dirac's formalism with the full configuration space \cite{11}. This is an important difference from the analysis given by \cite{7,7a} where this fact was not considered. 

The organization of this paper is as follows. In the next sections, we briefly discuss of two-dimensional gravity proposed by JT  model and the BF formulation. In Section IV, we show how to arrive at the complete structure of the constraints, and we find the Dirac brackets. In Section V we obtain the generator of the general gauge transformations with Castellani method.
In Section VI, we develop a complete analysis of the BF model in the JT symplectic formalism, and we find the full constraints, the symplectic matrix.  In Section VII, we obtained  the gauge symmetries of the theory with the Montani-Wotzasek MW formalism \cite{12a}. We work with the configuration space field as symplectic variables with the purpose of reproducing all the Dirac results, as we will see the generalized JT brackets and Dirac's brackets coincide to each other. In Section VIII finally, we present the results.

\section{Two dimensional gravity of Jackiw-Teitelboim model}
Jackiw and Teitelboim JT \footnote{The generalized of JT model through a function $f(\mu_{i},R)$, where f is a suitable polynomial function of $\mu_{i}$ and R has been proposed by Lichtzeir and Odintsov \cite{4a}} developed a model simple but nontrivial of General Relativity, where include, in addition to the space-time the metric $g_{\mu\nu}$, a scalar field $\Psi(x)$  \cite{3,4}. 

The JT model is described down the next action

\begin{eqnarray}
S_{JT}[g_{\mu\nu},\Psi]=\frac{1}{2}\int d^{2}x\sqrt{-g}\Psi(R-2\Lambda).
\label{1}
\end{eqnarray}

The Euler-Lagrange equations for JT model are given by

\begin{eqnarray}
\frac{\delta S{_{JT}}[g_{\mu\nu},\Psi]}{ \delta \Psi}&:& \qquad R-2\Lambda=0,  \\
\frac{\delta S{_{JT}}[g_{\mu\nu},\Psi]}{ \delta g_{\mu\nu}}&:& \qquad \nabla_{\mu}\nabla_{\nu}\Psi+\Lambda g_{\mu\nu}\Psi=0.
\label{2}
\end{eqnarray}

The first equations of motion refer to Einstein's equation (2), the parameter $\Lambda$ playing the role of the cosmological constant and the second equation (3) is the equation of motion for the scalar field \cite{5}.

\section{Jackiw-Teitelboim model in the BF formulation}

In n-dimensional spacetime, a BF theory\footnote{The symbol BF means that the lagrangian is made up of the exterior product of an (n-2) form B, with the curvature F of a connection A.} with gauge group $G$ involves two fields: a $G$ connection $A$, and a $\mathfrak{g}$-valued (n-2)-form $B$. In the absence of matter, the lagrangian is simply \cite{11a}
\begin{eqnarray}
L=Tr(B\wedge F).
\label{2b}
\end{eqnarray}

The BF theory of two dimensional gravity consist on the gauge connection 1-form $A$ and a scalar field $B$, also called background field ($B$-field), whose action is given by

\begin{eqnarray}
S_{BF}[B,A]=\int_{\mathcal{M}}Tr(B\wedge F),
\label{2c}
\end{eqnarray}

the trace is taken on the adjoint representation of $G$, and $F=dA+A\wedge A$ is the curvature of $A$. The corresponding Euler-Lagrange equations are
$F=0$ and $D_{A}B=0$, where the first equation simply say that the connections $A$ is flat, and the second equation say that $B$ is covariant constant-its covariant exterior derivative $D_{A}B$ vanishes.

It is well-know that geometric dynamics of JT model can be written as a BF theory in two dimensions \cite{Fu}, which is a generalization of Witten's work in the dimensions on Chern-Simons theory \cite{2,Fu}. The gauge group $G$ of two dimensional gravity is given by de Sitter or anti-de Sitter in Riemmanian or Lorentzian space-time.
Hence, the (A)dS algebra of the generators will satisfy the following commutation relations\footnote{The antisymmetric tensor $\epsilon_{IJ}$ is defined by $\epsilon_{01}=1$. The indices $I,J,...=0,1$ are lowered raised by the flat metric $\eta_{IJ}=diag(\sigma,1)$ or its inverse $\eta^{IJ}$, where $\sigma=\pm 1$ for Riemannian, resp. Lorentzian theory. For $\sigma=-1$ or 1, is the Lie algebra of $SO(3)$ or $SO(1,2)$ \cite{Fu, 7}.}

\begin{equation}
[\textbf{J}, \textbf{P}_{I} ]= \epsilon_{I}{^{J}}\textbf{P}_{J}, \quad \quad \quad  [ \textbf{P}_{I}, \textbf{P}_{J} ]= \epsilon_{IJ}\Lambda\textbf{J} ,
\label{5}
\end{equation}

where the operators $\textbf{P}_{I}$ and $\textbf{J}$ are the translation generators and the Lorentz boost generator.  \\
If we define generators of the (A)dS algebra as $\{T_{i}\}=\{T_{0},T_{1},T_{2}\}=\{\textbf{P}_{0},\textbf{P}_{1},\textbf{J}\}$, the Lie algebra can be written as\footnote{The completely antisymetric tensor $\epsilon_{ijl}$ is defined by $\epsilon_{012}=1$.}

\begin{eqnarray}
[T_{i},T_{j}]=f{_{ij}}^{k}T_{k}=\Lambda\epsilon_{ijl}\eta^{lk}T_{k},
\label{10}
\end{eqnarray}

and the metric $\eta_{ij}$ on the Lie algebra can be expressed as

\begin{eqnarray}
\eta_{ij}=diag(\Lambda\sigma,\Lambda,1).
\label{11}
\end{eqnarray}

Nonetheless, the presence of a non vanishing cosmological constant $\Lambda$ is necessary to ensure the non degeneracy of the Killing metric  and  to build a consistent gauge theory  \cite{Fu,11a}.

\section{ Hamiltonian dynamics of the Jackiw-Teitelboim model in the BF formulation}

The "BF" action (\ref{2c}) can be written in the form
\footnote{The notation "Tr"  represents a non-degenerate invariant bilinear form on the Lie algebra and $Tr(T_{i}T_{j})=\langle T_{i},T_{j}\rangle=\eta_{ij}$ plays the role of a metric on the Lie algebra.} \cite{Fu}

\begin{eqnarray}
S_{BF}[B,A]= \int Tr( B \wedge F ) \quad \rightarrow \quad S_{BF}[\phi,A]= \int Tr( \phi \wedge F ).
\label{14a}
\end{eqnarray}

The "B-field" ($B\rightarrow\phi=\phi^{i}T_{i}=: \varphi^{I}\textbf{P}_{I}+\Psi\textbf{J}$ ) of the theory is a 0-form  and $F=F^{i}T_{i}= F^{I}\textbf{P}_{I}+F^{2}\textbf{J}$ is the Yang-Mills curvature,
where $F^{I}=de^{I}+\omega{^{I}}_{J}\wedge e^{J}$ and $F^{2}=d\omega+\frac{\Lambda}{2}\epsilon_{IJ}e^{I}\wedge e^{J}$ represent the torsion and curvature 2-form of the zweibein field in the first order formalism\footnote{The values of the spacetime of the indices $\mu,\nu,...$ are label by t, x. The antisymmetric tensor $\epsilon^{\mu\nu}$ is defined by $\epsilon^{tx}=1$ \cite{7}.}.

The action (\ref{14a}) can be now rewriten as 

\begin{eqnarray}
S_{BF}[\phi,A]&=& \int Tr( \phi \wedge F ) =  Tr(T_{i}T_{j})\int \phi^{i}\wedge F^{j}, \nonumber \\
&=& \langle T_{i},T_{j}\rangle \int  \phi^{i}\wedge F^{j},\notag \ \\
&=& \frac{1}{2}\int d^{2}x\epsilon^{\mu\nu}\eta_{ij}\phi^{i} F^{j}{_{\mu\nu}}.
\label{14}
\end{eqnarray}

The variation of the action (\ref{14}) leads to the equations

\begin{eqnarray}
\frac{\delta S{_{BF}}[\phi,A]}{ \delta \phi_{i}}: \qquad  F^{i}=0, \qquad \frac{\delta S{_{BF}}[A,\phi]}{ \delta A^{i}}: \qquad D\phi_{i} =0,
\label{16a}
\end{eqnarray}

the equations (\ref{16a}) for two dimensional BF gravity are equivalent to equations JT model \cite{Fu}.

Therefore, in order to carry out the Hamiltonian analysis, we assume that space-time has the topological structure $M=\Sigma\times\Re$, where $\Sigma$ is a 1-dimensional manifold, representing the "space" and $\Re$ represents an evolution parameter. By performing the 1+1 decomposition, we can write the action (\ref{14}) as\footnote{The component fields are given by $\phi_{i}=(\varphi_{0},\varphi_{1},\Psi)$, $A^{i}_{x}=(e^{0}_{x},e^{1}_{x},\omega_{x})$, $A^{i}_{t}=(e^{0}_{t},e^{1}_{t},\omega_{t})$. $D_{x}\phi_{i}=\partial_{x}\phi_{i}+f{_{ij}}^{k}A^{j}_{x}\phi_{k}$.}

\begin{eqnarray}
S_{BF}[A,\phi]=\int d^{2}x(\phi_{i}\partial_{t}A^{i}_{x}+A^{i}_{t}D_{x}\phi_{i}).
\label{15a}
\end{eqnarray}

The definition of the momenta $(\Pi^{x}_{i},\Pi^{t}_{i},\Pi^{\phi}_{i})$ canonically conjugate to the configuration variables $( A^{i}_{x}, A^{i}_{t},\phi^{i})$ is given by
\begin{equation}
\Pi^{x}_{i}= \frac{\delta {\mathcal{L}} }{ \delta \dot{A}{^{i}}_{x} },  \qquad \Pi^{t}_{i}= \frac{\delta {\mathcal{L}} }{ \delta \dot{A}{{^i}}_{t} }, \qquad  \Pi^{\phi}_{i}= \frac{\delta {\mathcal{L}} }{ \delta \dot{\phi}{{^i}} } .
\label{16}
\end{equation}
The matrix elements of the Hessian is given by\footnote{Where $Q_{L}$ label the sets of variables $Q_{L}=\{A^{i}_{x},A^{i}_{t},\phi^{i}\}$.}
\begin{equation}
 H_{LM}=\frac{\partial^2{\mathcal{L}} }{\partial \dot{Q}_{L} \partial \dot{Q}_{M}}=0.
\label{17}
\end{equation}
Note that rank of the Hessian is zero, thus, we expect $9$ primary constraints. From the definition of the momenta $(\ref{16})$ we identify the following  $9$ primary constraints
\begin{eqnarray}
\Phi^{x}_{i}:= \Pi^{x}_{i}-\phi_{i} \approx 0 ,\qquad
\Phi^{t}_{i}:= \Pi^{t}_{i} \approx 0 , \qquad
\Phi^{\phi}_{i}:= \Pi^{\phi}_{i} \approx 0 .
\label{18}
\end{eqnarray}
The canonical Hamiltonian is given by
\begin{equation}
H_c= -\int dx A^{i}_{t}D_{x}\phi_{i},
\label{19}
\end{equation}

and the corresponding primary Hamiltonian $H_{P}$

\begin{equation}
H_P= H_c + \int dx \left[  \lambda^{i}_{x} \Phi^{x}_{i}+ \lambda^{i}_{t} \Phi^{t}_{i} + \lambda^{i}_{\phi} \Phi^{\phi}_{i} \right],
\label{20}
\end{equation}
where $ (\lambda^{i}_{x}, \lambda^{i}_{t}, \lambda^{i}_{\phi})$ are the corresponding Lagrange multipliers associated of these constraints $(\Phi^{x}_{i},\Phi^{t}_{i},\Phi^{\phi}_{i})$. The fundamental Poisson brackets of the theory  are determined by the commutation relations
\begin{eqnarray}
\{ A^{i}_{\mu}(x), \Pi^{\nu}_{j}(y) \}  &=& \delta^{i}_{j} \delta^{\nu}_{\mu} \delta(x-y), \nonumber \\
\{ \phi^{i}(x),\Pi^{\phi}_{j}(y) \}  &=& \delta^{i}_{j} \delta(x-y).
\label{21}
\end{eqnarray}

The next steps is to observer if there are more constraints, so that, we calculate the following 9$\times$9 matrix whose entries are the Poisson brackets among the constraints (\ref{18})

\begin{eqnarray}
\{ \Phi^{\mu}_{i} (x), \Phi^{\nu}_{j} (y) \} &=& 0, \nonumber \\
\{ \Phi^{x}_{i} (x), \Phi^{\phi}_{j} (y) \} &=& -\eta_{ij}\delta(x-y),
\label{22}
\end{eqnarray}

and expressed in matrix form, namely,

\begin{eqnarray}
W=\{\widetilde{\Phi}^{A}(x),\widetilde{\Phi}^{B}(y)\}=
\left(
 \begin{array}{ccc}
   0&0&-\eta_{ij}\delta(x-y)\\
   0&0&0\\
   \eta_{ij}\delta(x-y)&0&0\\
  \end{array}
\right),
\label{22a}
\end{eqnarray}

where, $ \widetilde{\Phi}^{A}=(\Phi^{x}_{i},\Phi^{t}_{i}, \Phi^{\phi}_{i})$. It is easy to see that this matrix has rank=6 and 3 null-vectors. By using these 3 null-vectors and the evolution of $\Phi^{t}_{i}$ produces the following  $3$ secondary constraints
\begin{eqnarray}
\dot{\Phi}^{t}_{i}&=& \{\Phi^{t}_{i} (x), {H}_{P} \} \approx 0 \quad \Rightarrow \quad \beta_{i}:= D_{x}\phi_{i}\approx 0,
\label{23}
\end{eqnarray}

and consistency conditions of $\Phi^{x}_{i}$ and $\phi_{i}$ leads to $6$ Lagrangian multipliers

\begin{eqnarray}
\dot{\Phi}^{x}_{i}&=&\{\Phi^{x}_{i}(x),H_{P}\}\approx0 \Rightarrow \lambda^{i}_{\phi}+f{_{ij}}^{k}A^{j}_{t}\phi_{k}\approx0, \nonumber \\
\dot{\Phi}^{\phi}_{i}&=&\{\Phi^{\phi}_{i}(x),H_{P}\}\approx0 \Rightarrow \lambda^{i}_{x}-D_{x}A^{i}_{t}\approx0.
\label{24}
\end{eqnarray}

Consistency conditions of the secondary constraints leads to no new constraints. Having found all constraints, we need to identify
from the primary and secondary constraints which corresponds to first and second class. In order to classify  the full set of constraints, we have to calculate the rank and the null-vectors of the  12$\times$ 12 matrix whose entries will be the Poisson brackets between primary and secondary constraints, this is
\begin{eqnarray}
\{\Phi^{x}_{i} (x),\Phi^{\phi}_{j} (y) \} &=& -\eta_{ij}\delta(x-y), \nonumber \\
\{\Phi^{x}_{i} (x),\beta_{j} (y)\} &=& f{_{ij}}^{k}\phi_{k}\delta(x-y) ,\nonumber \\
\{\Phi^{\phi}_{i} (x),\beta_{j}(y) \} &=& -\eta_{ij}\partial_{y}(x-y)-f_{ijl}A^{l}_{y}\delta(x-y).
\label{25}
\end{eqnarray}
This matrix has a vanishing determinant. After a long calculation, we found that this matrix has a rank=6 and 6 null vectors, thus, the theory presents a set of 6 first class constraints and 6 second class constraints. The structure of the first class constraints is obtained by means the null vectors, where, the null vectors of the matrix (\ref{25}) are  given by
\begin{eqnarray}
 V^{1}_{i}&=&(0,\delta^{j}_{i},0,0)\delta(x-y),\nonumber \\
 V^{2}_i&=&(-\delta^{j}_{i}\partial_{x}+f{^{j}}_{ik}A^{k}_{x},0,-f{_{i}}^{jk}\phi_{k},\delta^{j}_{i})\delta(x-y).
\label{25a}
\end{eqnarray}

In order to identified the following 6  first class constraints, we used the contraction of the null vectors (\ref{25a}) with the constraints (\ref{18}) and (\ref{23})
\begin{eqnarray}
\gamma_{i}&=& D_{x}\chi^{x}_{i}+D_{x}\phi_{i}-f{_{i}}^{jk}\phi_{k}\chi^{\phi}_{j}  \approx 0,  \nonumber \\
\gamma^{t}_{i}&=&   \Phi^{t}_{i}=\Pi^{t}_{i}  \approx 0,
\label{26}
\end{eqnarray}
and the following 6 second class constraints
\begin{eqnarray}
\chi^{x}_{i}&=& \Phi^{x}_{i}=\Pi^{x}_{i}-\phi_{i} \approx 0,  \nonumber \\
\chi^{\phi}_{i}&=& \Phi^{\phi}_{i}=\Pi^{\phi}_{i} \approx 0.
\label{27}
\end{eqnarray}

At this point, it is worth noting that these constraints (\ref{27}) have not been reported in the full phase space. As was pointed out at the introduction, it is mandatory to know the correct structure of the constraints on the full phase space in order to get complete information of the fundamental gauge transformation and the Dirac brackets. As we well know, the structure of constraints are related to gauge symmetries, besides, they have an important role  an important role on the formulation  canonical approaches of quantization \cite{8a}.

We now give the complete algebra among the constraints \footnote{ Where $f{_{ij}}^{k}=\Lambda\epsilon{_{ij}}^{k}$. } (\ref{26}) and (\ref{27})

\begin{eqnarray}
\{\gamma_{i} (x),\gamma_{j} (y) \} &=& f{_{ij}}^{k}\gamma_{k}\delta(x-y)\approx0, \nonumber \\
\{\gamma_{i} (x),\gamma^{t}_{j} (y) \} &=& 0, \nonumber \\
\{\gamma^{t}_{i} (x),\gamma^{t}_{j} (y) \} &=& 0, \nonumber \\
\{ \gamma_{i} (x), \chi^{\phi}_{j}(y) \} &=& f{_{ij}}^{k}\chi^{\phi}_{k}\delta(x-y)\approx0,\nonumber \\
\{\gamma_{i} (x),\chi^{y}_{j} (y) \} &=&f{_{ij}}^{k}\chi^{y}_{k}\delta(x-y)\approx0,\nonumber \\
\{\chi^{\phi}_{i} (x),\chi^{y}_{j} (y) \} &=&\eta_{ij}\delta(x-y),
\label{28}
\end{eqnarray}

where we can observe that the algebra of constraints (\ref{28}) is closed and is the local version of the  Lie algebra of the group A(dS). These constraints generate the A(dS) gauge transformation. Additionally, with all the information obtained until now, we can construct the Dirac brackets. For this aim, we shall use the matrix whose elements are only the Poisson brackets among second class constraints, namely $C{_{\alpha\beta}(x,y)}=\{\zeta^{\alpha}(x),\zeta^{\beta}(y)\}$, given by

\begin{eqnarray}
[C_{\alpha\beta}(x,y)]_{ij}=
\left(
 \begin{array}{cccc}
   0&1\\
   -1&0\\
  \end{array}
\right)\eta_{ij}\delta(x-y),
\label{29}
\end{eqnarray}

the inverse matrix $[C_{\alpha\beta}(x,y)]^{ij}$ is given by

\begin{eqnarray}
[C_{\alpha\beta}(x,y)]^{ij}=
\left(
 \begin{array}{cc}
   0&-1\\
   1&0\\
  \end{array}
\right)\eta^{ij}\delta(x-y).
\label{30}
\end{eqnarray}
The Dirac  brackets among two functionals $A$, $B$  is defined as

\begin{align}
\begin{aligned}
\{A(x),B(y)\}_{D}=\{A(x),B(y)\}_{P}-\int dudv\{A(x),\zeta^{\alpha}(u)\}C{^{-1}}_{\alpha\beta}(u,v)\{\zeta^{\beta}(v),B(y)\},
\label{31}
\end{aligned}
\end{align}

where $\{A(x),B(y)\}_{P}$ is the usual Poisson bracket between the functionals $A$, $B$ and $\zeta^{\alpha}=(\chi^{\phi}_{i},\chi^{x}_{i})$ is the set of second class constraints. By using (\ref{30}) and (\ref{31}), yields the following Dirac's brackets of the theory
\begin{eqnarray}
\{A^{i}_{x}(x),\phi_{j}(y)\}_{D}&=&\delta^{i}_{j}\delta(x-y),
\label{31a}
\end{eqnarray} 
\begin{eqnarray}
\{A^{i}_{t}(x),\Pi^{t}_{j}(y)\}_{D}&=&\delta^{i}_{j}\delta(x-y),
\label{31aa}
\end{eqnarray}
\begin{eqnarray}
\{A^{i}_{x}(x),\Pi^{x}_{j}(y)\}_{D}&=&\delta^{i}_{j}\delta(x-y),
\label{32}
\end{eqnarray}

we can observer that in BF model the fields $A^{i}_{x}$ and $\phi_{i}$ they are non-commutative.

We calculate the Dirac brackets among the first and second class constraints, and we have found that non trivial part of the Dirac Brackets is given by

\begin{eqnarray}
\{\gamma_{i}(x),\gamma_{j}(y)\}_{D}= f{_{ij}}^{k}(\gamma_{k}+f{_{k}}^{mn}\chi^{x}_{m}\chi^{\phi}_{n})\delta(x-y).
\end{eqnarray}

According to the Dirac formalism, the Dirac brackets among first class constraints must be square of second class constraints and linear of first class constraints \footnote{Quadratic terms in (32) may be present since the square of second class constraint is a first class one \cite{9, 6a}. }. Additionally, the Dirac Brackets amongst the second class constraints $\{\zeta^{\alpha}(x),\zeta^{\beta}(y)\}_{D}=0$, and with all other quantities turn out be zero \cite{8a}.

On the other hand, the identification complete of the constraints and Lagrange multipliers will allow us to identify the extended action.  By using the first class constraints (\ref{26}), the second class class constraints (\ref{27}), and the multipliers Lagrange multipliers (\ref{24}) we find that the extended action takes the form

\begin{eqnarray}
S_{E}[\phi^{i},\Pi^{\phi}_{i},A^{i}_{\mu},\Pi^{\mu}_{i},\lambda^{i},\lambda^{i}_{t},u^{i}_{x},u^{i}_{\phi}]&=&\int d^{2}x (\Pi^{\phi}_{i}\dot{\phi}^{i}+ \Pi^{\mu}_{i}\dot{A}^{i}_{\mu}-\mathcal{H}\nonumber \\ 
 &&-\lambda^{i}\gamma_{i}-\lambda^{i}_{t}\gamma^{t}_{i}-u^{i}_{x}\chi^{x}_{i}-u^{i}_{\phi}\chi^{\phi}_{i}),
\label{33a}
\end{eqnarray}

and
\begin{eqnarray}
\mathcal{H}=-A^{i}_{t}\gamma_{i}=-A^{i}_{t}(D_{x}\chi^{x}_{i}+D_{x}\phi_{i}-f{_{i}}^{jk}\phi_{k}\chi^{\phi}_{j}),
\label{33b}
\end{eqnarray}

where $\lambda^{i},\lambda^{i}_{t},u^{i}_{x},u^{i}_{\phi}$, are the Lagrange multipliers that enforce the first and second class constraints. We are to observable, by considering the second class constraints as strong equation, that the Hamiltonian (\ref{33b}) is reduced to the usual expression found in the literature \cite{7,7a}, which is defined on a reduced phase space context. From the extend action we can identify the extend Hamiltonian, which is given by

\begin{eqnarray}
H_{E}=\int dx (\mathcal{H}+\lambda^{i}\gamma_{i}+\lambda^{i}_{t}\gamma^{t}_{i}),
\end{eqnarray}

thus, the extended Hamiltonian is a linear combination of first-class constraints as expected.

The equations of motion obtained from the extend Hamiltonian and brackets Dirac are expressed by

\begin{eqnarray}
\dot{\phi}_{i}&=&\{\phi,H_{E}\}_{D}=f{_{ij}}^{k}(\lambda^{j}-A^{j}_{t})\phi_{k}, \nonumber \\
\dot{A}^{i}_{x}&=&\{A^{i}_{x},H_{E}\}_{D}=D_{x}(A^{i}_{t}-\lambda^{i}),\nonumber \\
\dot{A}^{i}_{t}&=&\{A^{i}_{t},H_{E}\}_{D}=\lambda^{i}_{t},\nonumber \\
\dot{\Pi}^{x}_{i}&=&\{\Pi^{x}_{i},H_{E}\}_{D}=f{_{ij}}^{k}(\lambda^{j}-A^{j}_{t})\Pi^{x}_{k},\nonumber \\
\dot{\Pi}^{t}_{i}&=&\{\Pi^{t}_{i},H_{E}\}_{D}=\gamma_{i}.
\end{eqnarray}

\section{Gauge generator}

We will calculate the Fundamental Gauge Transformation (FGT) defined on the full phase space. The construction of the FGT is based on the Castellani method and the gauge generators are given by first class \cite{11b}. According to the Castellani method, the gauge generator is given by

\begin{eqnarray}
G=\int_{\sum}\left[D{_{t}}\tau^{i}\gamma^{t}_{i}+\varepsilon^{i}\gamma_{i}\right].
\label{34}
\end{eqnarray}

By using the gauge generator, we obtain the following gauge transformation on the phase space

\begin{eqnarray}
\delta{_{0}} \phi_{i}&=&f{_{ij}}^{k}\varepsilon^{j}\phi_{k}, \nonumber \\
\delta{_{0}} A ^{i}_{x}&=&-D_{x}\varepsilon^{i}, \nonumber \\
\delta{_{0}} A^{i}_{t}&=& D_{t}\tau^{i}, \nonumber \\
\delta{_{0}} \Pi^{\phi}_{i}&=& f{_{ij}}^{k}\varepsilon^{j}\chi^{\phi}_{k},\nonumber \\
\delta{_{0}} \Pi^{x}_{i}&=& f{_{ij}}^{k}\varepsilon^{j}\chi^{x}_{k}+f{_{ij}}^{k}\varepsilon^{j}\phi_{k},\nonumber \\
\delta{_{0}} \Pi^{t}_{i}&=&f{_{i}}{^{j}}{_{k}}\tau^{k}\gamma^{t}_{j}.
\label{35}
\end{eqnarray}

We can see that FGT of BF model are given by (\ref{35}) and do not correspond to diffeomorphisms. Nevertheless, it is well known that a theory with background independence is diffeomorphisms covariant, and this symmetry can be obtained from the FGT. Hence, the diffeomorphisms must be found  by  redefining the gauge parameters as $\tau^{i}=-\varepsilon^{i}=v^{\rho}A^{i}_{\rho}$, and where $v$ is a vector field

\begin{eqnarray}
\delta{_{0}} \phi_{i}&=&-f{_{ij}}^{k}v^{\rho}A^{j}_{\rho}\phi_{k},\nonumber \\
\delta{_{0}} A ^{i}_{\mu}&=&D_{\mu}(v^{\rho}A^{i}_{\rho}),
\label{36}
\end{eqnarray}

and the gauge transformation of the fields takes the following form\footnote{$\mathfrak{L}$ represents the Lie derivative.}

\begin{eqnarray}
\phi' {_{i}}           &  \rightarrow  &  \phi_{i}+\mathfrak{L}{_{v}}\phi_{i}-v^{\rho}D_{\rho}\phi_{i}, \nonumber \\
A' {^{i}}_{\mu}     &  \rightarrow &   A^{i}_{\mu}+\mathfrak{L}{_{v}}A^{i}_{\mu}+v^{\rho}F^{i}_{\nu\rho}.
\label{37}
\end{eqnarray}

The expression for  diffeomorphisms are obtained (on shell) from the FGT as an internal symmetry of the theory. For other hand, the symmetries obtained in (\ref{35}), are related with Poincar\'e transformation. We can redefine the gauge parameters as $\tau^{i}=-\varepsilon^{i}=\theta^{i}+v^{\rho}A^{i}_{\rho}$ \cite{12}

\begin{eqnarray}
\delta_{0} A^{i}_{\mu}&=&D_{\mu}\theta^{i}+\partial_{\mu}v^{\rho}A^{i}_{\rho}+v^{\rho}\partial_{\rho}A^{i}_{\mu}+v^{\rho}F^{i}_{\mu\rho}=\delta_{PGT}A^{i}_{\mu}+v^{\rho}F^{i}_{\mu\rho},\nonumber \\
\delta_{0}\phi_{i}&=&-f{_{ij}}^{k}\theta^{j}\phi_{k}+v^{\rho}\partial_{\rho}\phi_{i}-v^{\rho}D_{\rho}\phi_{i}=\delta_{PGT}\phi_{i}-v^{\rho}D_{\rho}\phi_{i},
\label{38}
\end{eqnarray}

where

\begin{eqnarray}
\delta_{PGT}A^{i}_{\mu}&=&D_{\mu}\theta^{i}+\partial_{\mu}v^{\rho}A^{i}_{\rho}+v^{\rho}\partial_{\rho}A^{i}_{\mu},\nonumber \\
\delta_{PGT}\phi_{i}&=&-f{_{ij}}^{k}\theta^{j}\phi_{k}+v^{\rho}\partial_{\rho}\phi_{i}.
\label{39}
\end{eqnarray}

We can see that the gauge symmetries (\ref{38}) take back to the Poincar\'e symmetries up to terms proportional to the equations of motion (\ref{16a}).

\section{Faddev-Jackiw symplectic analysis for BF theory}
In this section, we focus now on the FJ method. In order to perform this aim, we observe that Lagrangian density of the action (\ref{15a}) can be written by a first-order Lagrangian like 

\begin{eqnarray}
\mathcal{L}(\xi, \dot{\xi})=\dot{\xi}^{(0)a}K^{(0)}_{a}(\xi)-V^{(0)}(\xi), \qquad a=1, 2, 3.  
\label{39a}
\end{eqnarray}

Using (\ref{39a}), we may rewrite (\ref{15a})

\begin{eqnarray}
\mathcal{L}^{(0)}=\phi_{i}\partial_{t}A^{i}_{x}+A^{i}_{t}D_{x}\phi_{i}=\phi_{i}\partial_{t}A^{i}_{x}- V^{(0)},
\label{40}
\end{eqnarray}

where the superscript $^{(0)}$ means initial Lagrangian and  $V^{(0)}=-A^{i}_{t}D_{x}\phi_{i}$ is called the symplectic  potential.

In the JT method, the Euler-Lagrange equations of motion are \cite{8b}
\begin{eqnarray}
f^{(0)}_{ab}(x,y)\dot{\xi}^{(0)b}(y)&=&\frac{\partial V^{(0)}(y)}{\partial\xi^{(0)a}(x)},  \\
f^{(0)}_{ab}(x,y)&=&\frac{\delta \mathrm{K}^{(0)}_{b}(y)}{\delta\xi^{(0)a}(x)}-\frac{\delta \mathrm{K}^{(0)}_{a}(x)}{\delta\xi^{(0)b}(y)},
\label{41}
\end{eqnarray}

where $f^{(0)}_{ab}$ is the symplectic matrix and with $\xi{^{(0)a}}$ representing a set of symplectic variables, $\mathrm{K}^{(0)}_{a}$ is called the canonical 1-form.
From expression (\ref{40}) we can identify the coefficients $\mathrm{K}^{(0)}_{a}(x)= \{\mathrm{K}^{(0)}_{1},\mathrm{K}^{(0)}_{2},\mathrm{K}^{(0)}_{3}\}= \{0,\phi_{i},0\}$.  In order to obtain all the Dirac results of previous section, we will use the configuration space $ \xi{^{(0)a}}(x)=\{\xi{^{(0)1}},\xi{^{(0)2}} , \xi{^{(0)3}}\} =\{\phi_{i},A^{i}_{x},A^{i}_{t}\}$ as symplectic variables  \cite{8b}. 
It is important to comment, that in JT framework we are free to choose  the symplectic variables,  we can choose the configuration variables or  the phase space variables.

The symplectic matrix  is given by

\begin{table}[ht]
\begin{eqnarray}
f^{(0)}_{ab}(x,y) = \left[ 
\begin{array}{l|cccc}
       & \phi_{j}    & A^{j}_{y}    & A^{i}_{t}     \\ \hline
\phi_{i}    & 0 & \delta^{i}_{j} & 0   \\
A^{i}_{x}    & -\delta^{j}_{i} & 0 & 0   \\
A^{i}_{t}    & 0 & 0 & 0 
\end{array}
\right]\delta(x-y).
\label{43}
\end{eqnarray}
\end{table}

The symplectic matrix $f^{(0)}_{ab}$ represents a $[9\times9]$ singular matrix. In JT scheme, this implies that the theory has constraints. In order to obtain these constraints, we calculate  the zero modes of the symplectic matrix (\ref{43}), in this case we have a zero mode, and is given by  $(v^{(0)a})_{1}^{T}=(0,0,v^{A^{i}_{t}})$, where $v^{A^{i}_{t}}$ is an arbitrary function. By multiplying the two sides of (45) by the zero modes $(v^{(0)a})_{1}^{T}$, the left-hand  side of (45) is 
 
\begin{eqnarray}
\int dx (v^{(0)a})_{1}^{T}(x)f^{(0)}_{ab}(x,y)=0.
\label{43a}
\end{eqnarray}
 Then, the right-hand side from (45) we can get the primary constraint as 
\begin{eqnarray}
\Omega^{(0)}_{i}&=&\int dx(v^{(0)a})^{T}(x)\frac{\delta}{\delta\xi^{(0)a}(x)}\int dy V^{(0)}(y), \nonumber \\
            &=& -\int dx v^{A^{i}_{t}}(x)\delta^{j}_{i}D_{x}\phi_{j},\nonumber  \\
&=& -v^{A^{i}_{t}}\delta^{j}_{i}D_{x}\phi_{j}=0,
    \label{44}
\end{eqnarray}

since $v^{A^{i}_{t}}$ is an arbitrary function, we obtain the following constraint
\begin{eqnarray}
\Omega^{(0)}_{i}=\delta^{j}_{i}D_{x}\phi_{j}=0,
\label{45}
\end{eqnarray}

this constraint is the secondary constraint found by means of Dirac's method in above section. In order to determine whether there are more constraints, we calculate the following  \cite{10a,10b, 10c, 10d}
\begin{eqnarray}
   F^{(1)}_{cb}\dot{\xi}^{b}=Z_{c}(\xi),
\label{46}
\end{eqnarray}

where
\begin{eqnarray}
Z_{c}(\xi)=
\left(
 \begin{array}{cccc}
   \frac{\partial V^{(0)}(\xi)}{\partial \xi^{a}}\\
   0\\
  \end{array}
\right),
\label{47}
\end{eqnarray}

and

\begin{table}[ht]
\begin{eqnarray}
F^{(1)}_{cb}(x,y) = \left(
 \begin{array}{ccc}
   f^{(0)}_{ab}\\
   \frac{\partial\Omega^{(0)}}{\partial\xi^{b}}\\
  \end{array}
\right)= \left[ 
\begin{array}{l|cccc}
       & \phi_{j}    & A^{j}_{y}    & A^{i}_{t}     \\ \hline
\phi_{i}    & 0 & \delta^{i}_{j} & 0   \\
A^{i}_{x}    & -\delta^{j}_{i} & 0 & 0   \\
A^{i}_{t}    & -\delta^{j}_{i} & 0 & 0   \\
\frac{\partial\Omega^{(0)}}{\partial\xi^{(0)b}}    &  \delta^{j}_{i}\partial_{x}-f{_{i}}{^{j}}_{k}A^{k}_{x} & f{_{ij}}^{k}\phi_{k} & 0 
\end{array}
\right]\delta(x-y).
\label{48}
\end{eqnarray}
\end{table}

We can observe that the matrix (\ref{48}) is  not a square matrix, nevertheless, note that this matrix has an independent mode given by $(v^{(1)})_{1}^{T}=(-f{_{ji}}^{k}\phi_{k}\delta(x-y),\delta^{i}_{j}\partial_{x}\delta(x-y)-f{_{j}}{^{i}}_{k}A^{k}_{a}\delta(x-y),v^{A_{t}},\delta^{i}_{j}\delta(x-y))$, this mode is used in order to obtain further  constraints. By means of the following expression \cite{10a, 10b, 10c, 10d}

\begin{equation}
(v^{(1)})_{c}^{T}Z_{c}=0,
\label{49}
\end{equation}
where $c=1$, we obtain that (\ref{49}) is an identity, thus, leads to no new constraints for the theory under study.\\
According to the JT symplectic, we will write a new Lagrangian, this is done by means of the  $A^{i}_{t}=\dot{\lambda}^{i}$ Lagrange multiplier  associated to that constraint $\Omega^{(0)}_{i} $, therefore, we can write the next symplectic Lagrangian

\begin{equation}
\mathcal{L}^{(1)}=\phi_{i}\dot{A}^{i}_{x}+ \Omega^{(0)}_{i}\dot{\lambda}^{i}-V^{(1)},
\label{50}
\end{equation}

where $V^{(1)}=V^{(0)}\mid_{\Omega^{(0)}_{i}=0}=0$, the symplectic potential vanish  reflecting the general covariance of the theory, just like it is present in  General Relativity. From the first-order Lagrangian (53), we can identify the next  symplectic variables $\xi{^{(1)a}}(x)=\{\phi_{i},A^{i}_{x},\lambda^{i}\}$ and the new coefficients of 1-forms $\mathrm{K}^{(1)}_{a}(x)=\{0,\phi_{i},\Omega^{(0)}_{i}\}$. Therefore, having considered this new information, we can obtain the following symplectic matrix

\begin{table}[ht]
\begin{eqnarray}
f^{(1)}_{ab}(x,y) = \left[ 
\begin{array}{l|cccc}
       & \phi_{j}    & A^{j}_{y}    & \lambda^{j}    \\ \hline
\phi_{i}    & 0 & \delta^{i}_{j} & \delta^{i}_{j}\partial_{y}+f{_{jk}}^{i} A^{k}_{y}   \\
A^{i}_{x}    & -\delta^{j}_{i} & 0 & -f{_{ij}}^{k}\phi_{k}   \\
\lambda^{i}    &  -\delta^{j}_{i}\partial_{x}-f{_{ik}}^{j}A^{k}_{x} & -f{_{ij}}^{k}\phi_{k} & 0 
\end{array}
\right]\delta(x-y),
\label{51}
\end{eqnarray}
\end{table}

where rows and columns follow the order $\phi_{i},A^{i}_{x},\lambda^{i}$.  The symplectic matrix $f^{(1)}_{ab}$ represents a $[9\times9]$ singular matrix. However, as discussed above there are not more constraints; the noninvertibility of (\ref{51}) implies that there is a gauge symmetry. If we want to invert the symplectic matrix, we choose the following gauge fixing
\begin{eqnarray}
A^{i}_{t}(x)&=&0,
\label{52}
\end{eqnarray}

according to the JT symplectic formalism, we have to introduce the gauge fixing as constraint by mean of lagrange multiplier $\sigma_{i}$.
Now, introducing this new information into (\ref{50}), leads to new symplectic Lagrangian

\begin{equation}
\mathcal{L}^{(2)}= \phi_{i}\dot{A}^{i}_{x}+(\Omega^{(0)}_{i}+\sigma_{i})\dot{\lambda}^{i},
\label{53}
\end{equation}

thus, we identify the following set of symplectic variables $\xi{^{(2)a}}(x)=\{\phi_{i},A^{i}_{x},\lambda^{i},\sigma_{i}\}$  and the symplectic 1-forms $\mathrm{K}^{(2)}_{a}(x)=\{0,\phi_{i},\Omega^{(0)}_{i}+\sigma_{i},0\}$. Furthermore, by using these symplectic variables we find that the symplectic matrix is given by

\begin{table}[ht]
\begin{eqnarray}
f^{(2)}_{ab}(x,y) = \left[ 
\begin{array}{l|cccc}
       & \phi_{j}    & A^{j}_{y}    & \lambda^{j} & \sigma_{j}   \\ \hline
\phi_{i}    & 0 & \delta^{i}_{j} & \delta^{i}_{j}\partial_{y}+f{_{jk}}^{i} A^{k}_{y} &0   \\
A^{i}_{x}    & -\delta^{j}_{i} & 0 & -f{_{ij}}^{k}\phi_{k} & 0  \\
\lambda^{i}    &  -\delta^{j}_{i}\partial_{x}-f{_{ik}}^{j}A^{k}_{x} & -f{_{ij}}^{k}\phi_{k} & 0 & -\delta^{j}_{i} \\
\sigma_{i}    & 0 & 0 & \delta^{i}_{j} & 0
\end{array}
\right]\delta(x-y).
\label{54}
\end{eqnarray}
\end{table}

The symplectic matrix $f^{(2)}_{ab}$ represents a $[12\times12]$ nonsingular matrix. After a long calculation, the inverse is given by

\begin{eqnarray}
[f^{(2)}_{ab}(x,y)]^{-1}=
\left(
 \begin{array}{cccccc}
   0&-\delta^{i}_{j}&0&-f{_{ji}}^{k}\phi_{k}\\
\delta^{j}_{i}&0&0&-\delta^{j}_{i}\partial_{y}-f{_{ik}}^{j}A^{k}_{y}\\
0&0&0&\delta^{j}_{i}\\
-f{_{ji}}^{k}\phi_{k}&\delta^{i}_{j}\partial_{x}+f{_{jk}}^{i}A^{k}_{x}&-\delta^{i}_{j}&0\\
  \end{array}
\right)\delta(x-y).
\label{55}
\end{eqnarray}
Therefore, from (\ref{55}) it is possible to identify the following JT generalized brackets by means of
\begin{eqnarray}
\{\xi_{i}^{(2)}(x),\xi_{j}^{(2)}(y)\}_{FJ}=[f^{(2)}_{ij}(x,y)]^{-1},
\end{eqnarray}
thus, the following brackets are identified
\begin{eqnarray}
\{A^{i}_{x}(x),\phi_{j}(y)\}_{FJ}&=&\delta^{i}_{j}\delta(x-y),\nonumber\\
\{\phi_{i}(x),\sigma_{j}(y)\}_{FJ}&=&-f{_{ji}}^{k}\phi_{k}\delta(x-y),\nonumber\\
\{A^{i}_{x}(x),\sigma_{j}(y)\}_{FJ}&=&-\delta^{i}_{j}\partial_{y}\delta(x-y)-f{^{i}}_{kj}A^{k}_{y}\delta(x-y),\nonumber\\
\{\lambda^{i}(x),\sigma_{j}(y)\}_{FJ}&=&\delta^{i}_{j}\delta(x-y).
\label{56}
\end{eqnarray}
It is important to comment,  that the generalized JT brackets  obtained from (\ref{55}) agreed with the brackets Dirac (\ref{31a}). In fact, if we make a redefinition of the fields introducing the momenta given by
\begin{eqnarray}
\Pi^{x}_{i}&=& \phi_{i},\nonumber \\ \Pi^{\phi}_{i}&=&0,
\label{57}
\end{eqnarray}
the generalized JT brackets (\ref{56}) take the form

\begin{eqnarray}
\{A^{i}_{x}(x),\phi_{j}(y)\}_{FJ}&=&\delta^{i}_{j}\delta(x-y),\nonumber \\
\{A^{i}_{x}(x),\Pi^{x}_{j}(y)\}_{FJ}&=&\delta^{i}_{j}\delta(x-y),
\label{58}
\end{eqnarray}

where we can observe that  (\ref{58}) coincide with  the full Dirac's brackets found in (\ref{31a}, \ref{32}) if we choose to impose the gauge $A^{i}_{t}=0$ in the Dirac's method.

As we have discussed earlier, in JT approach it is not necessary classify the constraints in first class or second class, since all the constraints are at the same footing. Therefore, we can perform the counting of physical degrees of freedom in the following form; there are $6$  dynamical variables $(\phi_{i}, A^{i}_{x})$ and $6$ constraints $(\Omega^{(0)}_{i} ,A^{i}_{t})$, therefore, the theory lacks of physical degrees of freedom.\\

\section{Gauge generator in the MW method}

Finally, we have calculated the gauge transformations of the theory, for this aim we calculate the mode of the matrix (\ref{51}), this mode is given by\footnote{In this context $[]^{T}$ means matrix transposition.}
\begin{eqnarray}
[W^{(1)a}]^{T}=(f{_{l}}^{ik}\phi_{k},\eta^{il}\partial_{x}+f^{li}{_{k}}A^{k}_{x},-\eta^{il})\delta(x-y).
\label{58a}
\end{eqnarray}
It can be seen that the zero-mode $(W^{(1)a})^{T}$ is the generator of the infinitesimal gauge symmetry on the constraint surface of the action (\ref{40}) and the infinitesimal gauge transformation of fields in compact notation is given by $\delta_{G}\xi^{(1)a}= \int dx [W^{(1)a}]^{T} \varepsilon  $ and $"\varepsilon "$ is a set of infinitesimal arbitrary  parameters.

From the above, we can see that the gauge transformation is therefore given by 
\begin{eqnarray}
\delta_{G} \xi^{(1)a}= (\delta_{G} \phi_{i}, \delta_{G} A^{i}_{x}, \delta_{G} \lambda^{i} )= \int dx (f{_{i}}^{jk}\phi_{k},\eta^{ji}\partial_{x}+f^{ij}{_{k}}A^{k}_{x},-\eta^{ji})\delta(x-y) \varepsilon_{j},
\end{eqnarray}
or, more explicitly 
\begin{eqnarray}
\delta_{G} \phi_{i}(x)=f{_{ij}}^{k}\phi_{k}\varepsilon^{j} ,\quad \delta_{G} A^{i}_{x}(x)=-D_{x}\varepsilon^{i}  ,\quad \delta_{G} \lambda^{i}(x)=-\varepsilon^{i}.
\end{eqnarray}

Now, coming back to the original variables\footnote{$\partial_{t}\delta \lambda^{i}=\delta A^{i}_{t}$.} 

\begin{eqnarray}
\delta_{G} \phi_{i}(x)=f{_{ij}}^{k}\phi_{k}\varepsilon^{j} ,\quad \delta_{G} A^{i}_{x}(x)=-D_{x}\varepsilon^{i}  ,\quad
\delta_{G} A^{i}_{t}(x)=-\partial_{t}\varepsilon^{i}= \partial_{t}\tau^{i}.
\label{59}
\end{eqnarray}

In this manner, by using the JT symplectic framework we have reproduced the first two components of gauge transformations reported in Dirac's method but the last component fails to do so. 

Montani and Wotzasek MW \cite{12a} developed one scheme of dealing with restrictions in the FJ method. Besides, they modified the FJ approach for obtained the symmetry transformations of action over all the configuration space, contrary to transformation (\ref{59}) generated by  JT formalism that only holds on the constraint surface. According to the MW method one should write the funcional variation of the corresponding Lagrangians\footnote{In term of matrix form $[\frac{\delta \mathcal{L}^{(1)}}{\delta \xi^{(1)a}}]^{T}=
\left(
\begin{array}{ccc}
\frac{\delta \mathcal{L}^{(1)}}{\delta \xi^{(1)1}} & \frac{\delta \mathcal{L}^{(1)}}{\delta \xi^{(1)2}} & \frac{\delta \mathcal{L}^{(1)}}{\delta \xi^{(1)3}}
\end{array} 
\right)$.} to zero.  Therefore,

\begin{eqnarray}
\frac{\delta \mathcal{L}^{(1)}}{\delta \xi^{(1)a}}= f^{(1)}_{ab}\dot{\xi}^{(1)b}-\frac{\partial V^{(1)}}{\partial \xi^{(1)a}}. 
\label{59a}
\end{eqnarray}
So, according to MW, on multiplying (\ref{59a}) by the zero-mode (\ref{58a}), we have

\begin{eqnarray}
[W^{(1)a}]^{T}\left(\frac{\delta \mathcal{L}^{(1)}}{\delta \xi^{(1)a}}= f^{(1)}_{ab}\dot{\xi}^{(1)b}-\frac{\partial V^{(1)}}{\partial \xi^{(1)a}}\right), 
\label{60}
\end{eqnarray}

from (\ref{60}), we obtained the following equations\footnote{Where $D^{i}_{jy}=\delta^{i}_{j}\partial_{y}+f{_{jk}}^{i} A^{k}_{y}$ and $D^{j}_{ix}=\delta^{j}_{i}\partial_{x}+f{_{ik}}^{j}A^{k}_{x}$.}

\begin{eqnarray}
(f{_{l}}^{ik}\phi_{k},\eta^{il}\partial_{x}+f^{li}{_{k}}A^{k}_{x},-\eta^{il})\left(
\left(
\begin{array}{c}
\frac{\delta \mathcal{L}^{(1)}}{\delta \phi_{i}} \\ 
\frac{\delta \mathcal{L}^{(1)}}{\delta A^{i}_{x}} \\ 
\frac{\delta \mathcal{L}^{(1)}}{\delta \lambda^{i}}
\end{array}
\right)= \left[ 
\begin{array}{cccc}
      
 0 & \delta^{i}_{j} & D^{i}_{jy}   \\
 -\delta^{j}_{i} & 0 & -f{_{ij}}^{k}\phi_{k}   \\
 -D^{j}_{ix} & -f{_{ij}}^{k}\phi_{k} & 0 
\end{array}
\right]\left(
\begin{array}{c}
\dot{ \phi}_{j} \\ 
\dot{A}^{j}_{x} \\ 
\dot{\lambda}^{j}
\end{array}
\right)\right),
\label{60a}
\end{eqnarray}

from eq.(\ref{60a}), one finds  the GT over all the configuration space. Using this equations, the GT  can be written as

\begin{eqnarray}
f{_{l}}^{ik}\phi_{k}\frac{\delta \mathcal{L}^{(1)}}{\delta \phi_{i}}+(\eta^{il}\partial_{x}+f^{li}{_{k}}A^{k}_{x})\frac{\delta \mathcal{L}^{(1)}}{\delta A^{i}_{x}}-\eta^{il}\frac{\delta \mathcal{L}^{(1)}}{\delta \lambda^{i}}&=&f{_{l}}^{ik}\phi_{k}(A^{i}_{x}-D_{x}\dot{\lambda}^{i})+(\eta^{il}\partial_{x}+f^{li}{_{k}}A^{k}_{x})(-\dot{\phi}_{i}- f_{iqp}\phi^{p}\dot{\lambda}^{q})+\nonumber \\ 
 &&(-\eta^{il})(-\dot{\Omega}^{(0)}_{i}). \nonumber 
\end{eqnarray}

Now, coming back to the original variables

\begin{eqnarray}
\dot{\lambda}^{i}&\rightarrow& A^{i}_{t},\nonumber \\
\frac{\delta \mathcal{L}^{(1)}}{\delta \lambda^{i}}&\rightarrow& \frac{\delta \mathcal{L}^{(1)}}{\delta A^{i}_{t}}\partial_{t}, \nonumber \\
\frac{\delta \mathcal{L}^{(1)}}{\delta \phi_{i}}&=& \frac{\delta \mathcal{L}^{(0)}}{\delta \phi_{i}},\nonumber \\
\frac{\delta \mathcal{L}^{(1)}}{\delta A^{i}_{x}}&=&\frac{\delta \mathcal{L}^{(0)}}{\delta A^{i}_{x}},
\label{60aa}
\end{eqnarray}

and implementing the equations of motion for the gauge fields

\begin{eqnarray}
\frac{\delta \mathcal{L}^{(0)}}{A^{i}_{t}}=\Omega^{(0)}_{i},
\label{61a}
\end{eqnarray}

we obtain\footnote{$\eta^{il}D_{t}=\eta^{il}\partial_{t}+f{^{l}}{_{n}}^{i}A^{n}_{t}$.}

\begin{eqnarray}
(-1)\eta^{il}D_{t}\varepsilon_{l}\frac{\partial \mathcal{L}^{(0)}}{\partial A^{i}_{t}} + (-1)\eta^{il}D_{x}\varepsilon_{l}\bigg(\frac{\partial \mathcal{L}^{(0)}}{\partial A^{i}_{x}}-\partial_{t}\frac{\partial \mathcal{L}^{(0)}}{\partial\dot{A}^{i}_{x}}\bigg)+f{_{i}}^{lk}\phi_{k}\varepsilon_{l}\frac{\partial \mathcal{L}^{(0)}}{\partial \phi_{i}}=0.
\end{eqnarray}

Then, we can found that the gauge field transformations are given by 

\begin{eqnarray}
\delta_{G} A^{i}_{t}(x)=-D_{t}\varepsilon^{i}= D_{t}\tau^{i},\quad \delta_{G} A^{i}_{x}(x)=-D_{x}\varepsilon^{i}  ,\quad
\delta_{G} \phi_{i}(x)=f{_{ij}}^{k}\phi_{k}\varepsilon^{j}.
\label{62}
\end{eqnarray}
This means that the transformations will be a gauge symmetry over all the configuration space. We found the same transformations as obtained by using Dirac approach. 

\section{ Conclusions and prospects}

In this work, we analyse the BF model of JT theory from point of view of the Dirac formalism and FJ symplectic  method for constrained systems. In Dirac formalism, we find the first and
the second-class constraints, and we can see that  $S_{BF}$ action for gravity in two dimensions is devoid of degrees of freedom, this is clearly in
concordance with the fact this theory contain  $18$ canonical variables $(\phi_{i}, A^{i}_{\mu}, \Pi^{\phi}_{i},\Pi^{\mu}_{i})$, $6$ first class constraints $(\gamma_{i}, \gamma^{t}_{i})$ and $6$ second class constraints $(\chi^{\phi}_{i}, \chi^{x}_{i})$, as a result ones get zero degrees of freedom $18-2(6)-6=0$,  consequently, the theory is topological. Furthermore, in this approach we have obtained the fundamental gauge structure as well the  algebra between the constraints, and it has been shown that the set of constraints form a closed algebra (AdS) (\ref{28}). Besides, by defining the gauge parameters, diffeomorphisms and Poincar\'e symmetries can be obtained from the fundamental gauge symmetry. On the other hand, considering the second class constraints as strong equality, the results is reduced to the usual expression found in the literature \cite{7,7a}, which is defined on a reduced phase space context.
These results obtained can be compared with those calculate by means of the Hamilton-Jacobi \cite{8} formalism, where, we have no distinction between the  first and the second- class constraints. 

In the JT symplectic approach we have a set of the constraints of the theory are at the same footing and generally leads to a less number of constraints that the Dirac formalism, and this fact allows that the JT symplectic method is more convenient to perform. Moreover, we have showed that the generalized JT brackets and the Dirac's ones coincide to each other. Besides, we have obtained that the number of physical degrees of freedom is the same as the one obtained from the  Dirac formalism. On the other hand, was obtained the  gauge symmetry over all the configuration space by using the MW algorithm.  
In this manner, we have reproduced all relevant Dirac's results by working with  JT symplectic, in particular we can see that JT symplectic method is more economical when it is compared with the Dirac formalism.

We finish this paper with some comments, as discussed above, in the JT symplectic framework it is not necessary to classify the constraints in second class or first class as in Dirac's method is done, consequently, the algebraic operations that involving constraints analysis are shortened. This fact allows that the JT symplectic  method is more convenient to develop. In this sense, we can carry out the analysis to other models of 2D gravity. The action (\ref{70}) is an alternative model reproducing Einstein's equations with a cosmological constant and dynamical torsion \cite{13, 13a}

\begin{eqnarray}
S[e^{I}_{\mu},\omega_{\mu}]=\int d^{2}x e\bigg(\frac{1}{16\alpha}R^{IJ}_{\mu\nu}R^{\mu\nu}_{IJ}-\frac{1}{8\beta}T^{I}_{\mu\nu}T^{\mu\nu}_{I}-\Lambda \bigg),
\label{70}
\end{eqnarray}

at the same time, the action (\ref{70}) contains solutions with constant curvature and zero torsion, it also includes several other 2D gravity models \cite{3, 4, Fu} and this is of particular interest for investigations of the quantum structure of gravity. The Hamiltonian analysis of the model (\ref{70}) has been developed in \cite{14} and its canonical quantization in \cite{15}. On the other hand, we can find in the literature \cite{13a} that the model (\ref{70}) in the region $e=det (e^{I}_{\mu})\neq 0$ can be written as a gauge theory based on the quadratic extension of the Poincar\'e algebra and can be rewritten as

\begin{eqnarray}
\tilde{S}[e^{I}_{\mu},\omega_{\mu},\varphi,\varphi_{I}]=\int d^{2}x [\frac{1}{2}\varepsilon^{\mu\nu}(\varphi R_{\mu\nu}+\varphi_{I}T^{I}_{\mu\nu})-e(\alpha\varphi^{2}+\beta\varphi_{I}\varphi^{I}+\Lambda)].
\label{71}
\end{eqnarray}

The Hamiltonian analysis on the full phase space of the action (\ref{71}) has not been reported, and the complete structure of the constraints is unknown. As previously discussed, in some cases, implementing the Dirac algorithm is large and tedious task, hence, it is necessary to use alternative formulations that could give us a complete canonical description of the theory. In this sense, we will utilize the JT symplectic formalism to study the action (\ref{71}) in forthcoming works \cite{16}.

\section*{Acknowledgments}

J.M.C welcomes the support of the Universidad Juárez Autónoma de Tabasco for providing a suitable work environment while this research was carried out. J.M.C. thanks CONACYT for the support through a grant for postdoctoral studies under grant No. 3873825.


\end{document}